\documentclass[reqno]{amsart}

\usepackage{amssymb,epsfig,graphics}
\usepackage{amsmath}
\usepackage{amscd}
\usepackage{verbatim}
\input xypic

\theoremstyle{definition}

\newtheorem{example}{Example}

\def\C{\mathbb C}
\def\R{\mathbb R}
\def\Z{\mathbb Z}

\def\r{\rangle}
\def\l{\langle}

\def\c{\cdot}

\begin{document}

\title[Branching of $A_n$ orbits]
{Branching rules for the Weyl group orbits\\ of the Lie algebra $A_n$}

\author{M. Larouche, M. Nesterenko$^\dag$, and J. Patera}
\address{Centre de recherches math\'ematiques,
         Universit\'e de Montr\'eal,
         C.P.~6128 Centre-ville,
         Montr\'eal, H3C\,3J7, Qu\'ebec, Canada;
         $^\dag$~Institute of mathematics of NAS of Ukraine,
         3, Tereshchenkivs'ka str, Kyiv-4, 04216, Ukraine}
\email{patera@crm.umontreal.ca, larouche@dms.umontreal.ca, maryna@imath.kiev.ua}

\date{\today}
 \begin{abstract}
The orbits of Weyl groups $W(A_n)$ of simple $A_n$ type Lie algebras
are reduced to the union of orbits of the Weyl groups of
maximal reductive subalgebras of $A_n$. Matrices transforming
points of the orbits of $W(A_n)$ into points of
subalgebra orbits are listed for all cases $n\leq8$ and for the infinite
series of algebra-subalgebra pairs $A_n\supset A_{n-k-1}\times
A_k\times U_1$, $A_{2n}\supset B_n$, $A_{2n-1}\supset C_n$,
$A_{2n-1}\supset D_n$. Numerous special cases and examples are
shown.
\end{abstract}

\maketitle

\section{Introduction}
Finite groups generated by reflections in an $n$-dimensional real
Euclidean space~$\R^n$ are commonly known as finite Coxeter groups
\cite{H1,H}. Finite Coxeter groups are split into two classes:
crystallographic and non-crystallographic groups. Crystallographic
groups are often referred to as Weyl groups of semisimple Lie groups
or Lie algebras. They are symmetry groups of some lattices in $\R^n$.
There are four infinite series (as to the admissible values of rank $n$)
of such groups, and five isolated exceptional groups of ranks 2, 4, 6, 7,
and 8. Non-crystallographic finite Coxeter groups are the symmetry groups
of regular $2D$ polygons (the dihedral groups), and two exceptional groups,
for $n{=}3$ (the icosahedral group of order 120) and $n{=}4$,
which is of order $120^2$.

We consider the orbits of the Weyl groups $W(A_n)$ of the simple Lie
algebras of type $A_n$, $n\geq1$, equivalently the Weyl groups of
the simple Lie group  $SL(n{+}1,\C)$, or of its compact real form
$SU(n{+}1)$. The order of such a Weyl group is $(n{+}1)!$. An orbit
of $W(A_n)$ is a set of distinct points in $\R^n$, obtained from a
chosen single (seed) point, say $\lambda\in\R^n$, by application of
$W(A_n)$ to $\lambda$. Hence, an orbit $W_\lambda$ of
$W(A_n)$ contains at most $(n{+}1)!$ points. The points of
$W_\lambda$ are equidistant from the origin. It should be noted
that the group $W(A_n)$ is isomorphic to the permutation group
of $n{+}1$ elements. Although we make no use of this fact here, it
reveals a rather different perspective on our problem \cite{NPT}.

Geometrically, points of the same orbit can be seen as vertices of a
convex polytope generated from $\lambda$. There is a method for
counting and describing the faces of all dimensions of such polytopes in
the real Euclidean space $\R^n$. It uses an easy recursive
decoration of the corresponding Coxeter-Dynkin diagrams \cite{CKPS}.

Weyl group orbits are closely related to weight systems of finite
irreducible representations of corresponding Lie algebras. More
precisely, the weight system is a union of several Weyl group
orbits. Which orbits are composed into a particular weight system is
in principle known. An efficient algorithm for the computation exists
\cite{BMP}. The representations are finding innumerable applications
in science. Very often, such applications can be carried through just
by our knowledge of the corresponding weight system. It is
conceivable that some of the applications would find interesting new
possibilities when working with individual orbits only.

The list of possible reductions of $W(A_n)$ orbits is a result of a
major classification problem solved more than half a century ago,
when the maximal reductive subalgebras of simple Lie algebras, in
particular of $A_n$, were determined \cite{BdeS,Dynkin}. We exploit
that classification without further reference to it.

In this paper, we consider orbits of $W(A_n)$ and their reduction to
orbits of the Weyl groups of maximal reductive subalgebras of $A_n$.
In the physics literature, a similar task \cite{McP} is often called
computation of branching rules. We will consider two types of maximal
reductive subalgebras, maximal reductive subalgebras that are
not semisimple \cite{BdeS}, and subalgebras that are maximal among
reductive subalgebras, but which are in fact semisimple. Thus the second type of subalgebras are obtained from the list
of \cite{Dynkin} by eliminating semisimple subalgebras that
are part of the reductive subalgebras classified in \cite{BdeS}.

The present paper can be understood as a continuation of \cite{HLP},
where the orbits are seen as elements of a ring of reflection
generated polytopes in $\R^n$. In that paper, the main problem was to reduce
products of Weyl group orbits/polytopes into a sum of Weyl group
orbits. Here, our problem is to transform/reduce/branch each
polytope/orbit into a sum of concentric polytopes with lower
symmetry, and often also with lower dimension.

Until recently, $W$-orbits were used as an efficient computational
tool, particularly for large-scale computations (see for example
\cite{BMP,GP,MP,MP1} and references therein). Their appreciation as
point sets defining families of $W$-invariant special functions of
$n$ variables is relatively recent \cite{KP1,KP2,P}. Other possible
applications could include an unusual twist of some symmetry
breaking problems in physics, where, rather than breaking down
weight systems of representations, one would break
each orbit independently.

The main advantage of the projection matrices method
is the uniformity of its application as to the
different algebra-subalgebra pairs, which makes it particularly
amenable to computer implementation. Thus in \cite{PSan}, branching
rules for representations of dimension up to 5000 were calculated
for all simple Lie algebras of rank up to 8 and for all their
maximal semisimple subalgebras. Corresponding projection matrices
were presented as a computational tool only later in \cite{MPS}.
Subsequently, the tables \cite{McP} were also based on their
exploitation.

Particular Weyl group orbit reduction has undoubtedly been addressed
on many occasions in the literature. As a separate subject of
interest, orbit branching rules seem to have been first found in \cite{MPR}, where they
are used for reduction of many representations as well as orbits of
the five exceptional simple Lie algebras. The corresponding
projection matrices are shown there too. In \cite{GPS}, several
generating functions for the reduction problem were derived. It is a
very efficient method, in that it solves the problem for all orbits
at once. Unfortunately, for each algebra-subalgebra pair, a new
generating function needs to be derived. An independent original
approach to orbit-orbit branching rules can be found in \cite{ST1,ST2},
in which essentially combinatorial algorithms are developed for specific
series of algebra-subalgebra pairs. For $A_n$, an algorithm for the
equal rank subalgebra series of cases can be found there. It should
be compared with subsection 4.3 of this paper.

Our problem in this paper is closely related to the computation of
branching rules for irreducible finite dimensional representations
of simple Lie algebras (equivalently, to branching rules for weight
systems of representations). Theoretically, such problems need to
be solved while describing symmetry breaking in some physical
systems. Practically, the orbit branching rules problem needs to be
solved whenever a large-scale computation of branching rules for
representations is undertaken. The similarity of the two problems is
in the transformation of orbit points (weights) that takes place in
both cases. However, there are practically important differences
between the two problems. The orbit branching rules are less
constrained than those for representations. Some of the differences
were already pointed out in \cite{HLP}. Here we underline just two:
\smallskip\newline
(i) \ \ While weight systems grow without limits, the larger
the representation one has to work with, the orbit size (the number of points
in an orbit) is always bounded by the order of the corresponding
Weyl group. Without limits, only the distance of the orbit
points from the origin can grow, but not their number. A weight system of a
representation is a union of several $W$-orbits. The higher the
representation, the more orbits it is comprised of. In general, to
determine the orbits that form the weight system of a representation
(equivalently, to compute dominant weights multiplicities in a
representation) is often a difficult and laborious task (see
\cite{BMP} and references therein). Therefore, any large-scale
computation with representations practically imposes the need to
break a large problem for the weight system, into a series of much
smaller ones for individual orbits. Computation of branching rules for the representations is one such problem, decomposition of products
of representations into the direct sum of irreducible
representations is another problem, which often needs to be carried out for
relatively large representations, and which is solved entirely using
the weight systems, see for example \cite{GP}.
\smallskip\newline
(ii) \ A point of a weight system of a representation necessarily
belongs to a weight lattice of the Lie algebra. Its coordinates are
integers in a suitably chosen basis of $\R^n$, so are the points of
orbits after reduction. When we work with an individual orbit,
we are free to choose the orbit, that is, the seed point $\lambda$,
anywhere in $\R^n$, as close or as distant from the
origin or from any other lattice point as one desires. After the
reduction, some orbits can be very close, while some are far apart.
Examples of such effects are shown in the Concluding Remarks
of~\cite{HLP}. The flexibility thus achieved needs yet to be
exploited.

The branching rules for $W(A_n)\rightarrow W(L),$ where $L$ is a
maximal reductive subalgebra of $A_n$, is a linear transformation
between Euclidean spaces $\R^n\rightarrow\R^m,$ where $m$ is the
rank of $L$. The branching rules are unique, unlike transformations
of individual orbit points, which depend on the relative choice of bases.
In this paper, we provide the linear transformation in the form of an
$n\times m$ matrix, the `projection matrix'. A suitable choice of
bases allows one to obtain integer matrix elements in all the
projection matrices listed here. Note that we use Dynkin notations
and numberings for roots, weights and diagrams.


\section{Preliminaries}

The general strategy of our approach can be described as follows.

Consider the pair $L\supset L'$ of Lie algebras of ranks $n$ and $m$
respectively, where $L$ is simple and $L'$ is maximal reductive. In
principle, the orbit reduction problem for the pair $W(L)\supset
W(L')$ is solved  when the $n\times m$ matrix $P$ is found, with the
property that points of any orbit of $W(L)$ are
transformed/projected by $P$ into points of the corresponding
orbits of $W(L')$. Computation of the branching rule for a specific
orbit amounts to applying $P$ to the points of the orbit, and to
sorting out the projected points according to the orbits of $W(L')$.

This task requires that one be able to calculate the
points of any orbit of the Weyl group of any semisimple Lie algebra
encountered here. There is a standard method, but we refrain from
describing it here once again. Instead
we refer to \cite{HLP}, the immediate predecessor of this paper, wherein
all orbit points are given relative to the so called
$\omega$-basis. Geometric relations between the basis vectors are
described by the matrix $(\l\omega_j,\omega_k\r)$ of scalar products
of the basis vectors. The matrices are found in \cite{BMP} under the
name `quadratic form matrices' for all simple Lie algebras.

The Weyl group of the one-parameter Lie algebra $U_1$ is trivial,
consisting of the identity element only. This algebra is present in
reductive non-semisimple Lie algebras. Its irreducible
representations are all 1-dimensional, hence its orbits consist of
one element. They are labeled by integers. The symbol $(k)$ may
stand for either the orbit $\{k,-k\}$ of $A_1$, or for the
$U_1$-orbit of one point $\{k\}$. Distinction should be made
from the context. For example, the orbit $(p)(q)$, where
$p\in\Z^{>0},\ q\in\Z$, of $W(A_1\times U_1)$, has two elements,
$\{(p)(q),(-p)(q)\}$.

All orbits of $W(A_n)$ have the following symmetry. For each point
$(a_1,a_2,\dots, a_n)$ that belongs to an orbit, the point
$(-a_n,-a_{n-1},\dots, -a_2,-a_1)$ also belongs to the same orbit. We say
that the orbits of $W(A_n)$ in the following pair are
contragredient:
\begin{gather*}
(q_1,q_2,\dots,q_{n-1},q_n),\qquad
(q_n,q_{n-1},\dots,q_2,q_1),\qquad
   q_j\geq 0\ \text{ for all }\ j.
\end{gather*}
Branching rules for contragredient orbits are closely related. They
either coincide, or one can be obtained from the other by
interchanging $q_k\ \leftrightarrow\ q_{n-k}$ components of the
dominant points. We list only one such pair of branching rules.

It is known that $W(A_n)$ orbits $(1,0,\dots,0)$, $(0,1,0,\dots,0)$,
$\cdots$, $(0,\dots,0,1)$ for $n\geq1$ constitute the entire
weight system of the corresponding irreducible representation. If no other orbits are involved in a branching rule, that rule coincides with the branching rule for representations.

The number of points in a Weyl group orbit, labeled by its unique
dominant weight $(a_1,a_2,\dots,a_n)$, is determined by the $a_j$'s
that are strictly positive. In orbits encountered in representation
theory, we have $a_j\in\Z^{\geq0}$. Since we are considering a more
general setup, we need require only $a_j\in\R^{\geq0}$. If all
$a_j$'s are strictly positive, the orbit of $W(A_n)$ contains
$(n{+}1)!$ points.

For simplicity of notation we subsequently identify cases by
algebra-subalgebra symbols rather than by corresponding Weyl
groups. In particular, we speak of an orbit of $A_k$ rather than of
an orbit of $W(A_k)$.

\section{Construction of projection matrices}

The projection matrix $P$ for a given pair $L\supset L'$ of Lie algebras
is calculated from one known branching rule. The classification of subalgebras amounts precisely to providing that branching rule. Usually the
branching rule is given for the lowest dimensional representation.
Then the matrix is obtained using the weight systems of the involved
representations.

First, make a suitable (lexicographical) ordering of the weights of
$L$ and ~$L'$. Then associate the weights on both sides
one-by-one according to the chosen order. The matrix is obtained from
requiring that each weight of $L$ be transformed to its associate
weight of~$L'$.
\begin{example}\

Consider the case of $A_3\supset C_2$ of subsection 5.2. The lowest
orbit of $A_3$ contains 4 points.
The lowest orbit of $C_2$ also contains 4 points. More
precisely, there are two 4-point orbits of $A_3$ and two such
orbits of $C_2$. Either of the two $A_3$ orbits can be used for
setting up the projection matrix. The two orbits of $C_2$ with dominant weights
$(1,0)$ and $(0,1)$ are different, being related to simple roots of
different length. We take the $A_3$~orbit of the dominant
point $(1,0,0)$ and project it onto the $C_2$~orbit of the point $(1,0)$. (See the second option in the last item of Concluding Remarks below.)
\begin{gather*}
(1,0,0)\mapsto (1,0),\qquad (\!{-}1,1,0)\mapsto (\!{-}1,1),\\
\!\!\!(0,\!{-}1,1)\mapsto (1,\!{-}1),\quad\; (0,0,\!{-}1)\mapsto (\!{-}1,0).
\end{gather*}

Writing the points as column matrices, the projection
matrix of subsection 5.2 is obtained from the first three. Proceeding
one column at a time, we have
\begin{alignat*}{3}
\left(\begin{smallmatrix} 1 & * & * \\ 0 & * & * \end{smallmatrix}\right)
           \left(\begin{smallmatrix} 1 \\ 0 \\ 0 \end{smallmatrix}\right)
           &=\left(\begin{smallmatrix} 1 \\ 0  \end{smallmatrix}\right),&\qquad
\left(\begin{smallmatrix} 1 & 0 & * \\ 0 & 1 & * \end{smallmatrix}\right)
           \left(\begin{smallmatrix}\!\!{-}1 \\ 1 \\ 0 \end{smallmatrix}\right)
           &=\left(\begin{smallmatrix}\!\!{-}1 \\ 1  \end{smallmatrix}\right),&\qquad
\left(\begin{smallmatrix} 1 & 0 & 1 \\ 0 &1 &0\end{smallmatrix}\right)
           \left(\begin{smallmatrix} 0 \\ \!\!{-}1 \\ 1 \end{smallmatrix}\right)
           &=\left(\begin{smallmatrix} 1 \\ \!\!{-}1  \end{smallmatrix}\right).
\end{alignat*}
Here, stars denote the entries that are still to be determined.
The matrix ${\mbox {\it P}{=}\left(\!\begin{smallmatrix} 1 & 0 & 1 \\ 0 &1 &0\end{smallmatrix}\!\right)}$
then automatically transforms the fourth point $(0,0,-1)$
of the $A_3$ orbit as required. This matrix can be used for
projecting points of any $A_3$ orbit.

\end{example}

\section{Equidimensional orbit branching rules}

All orbits $W(A_n)$ are $n$-dimensional except for the trivial one
$\lambda=0$, which consists of one point, the origin. Points
can be seen as vertices of a polytope in $\R^n$ \cite{HLP}.
Reduction to orbits of the same dimension happens when reduced
orbits have the symmetry of $W(A_r\times A_s\times U_1),$ where
$r{+}s{+}1=n$. Clearly, we need to consider only the cases $r\geq s$.
Geometrically, the orbit points are not displaced in this case; rather, they
are relabeled by the coordinates given in the standard
basis of the subgroup.

In this section, we first consider the lowest special cases in part
as transparent illustration and in part because they are
most frequently encountered in physics applications. Lastly, we
consider the infinite series of cases $1\leq n<\infty$ for all
possible values of the rank $n$:  $W(A_n)\rightarrow
W(A_{n-k-1}\times A_{k}\times U_1)$, $0\leq k\leq[\tfrac{n-1}2]$,
where $[\tfrac{n-1}2]$ is the integer part of $\tfrac{n-1}2$.

\subsection{Orbit branching rules for
$A_n\supset A_{n-1}\times U_1$}\

\subsubsection{$A_1\supset U_1$}\

The lowest example is trivial. The Weyl group of $A_1$ has two
elements; the Weyl group of $U_1$ is just the identity
transformation. An orbit $\{p,-p\}$ of $A_1$ reduces to two orbits
of $U_1$:
\begin{gather*}
(p)\supset (p){+}(-p),\qquad p\in\R^{>0}.
\end{gather*}
The reduction is accomplished by applying the $1\times1$ projection
matrix~$P=(1)$ to each element of the $A_1$ orbit.

\subsubsection{$A_2\supset A_1\times U_1$}\

The second lowest example is often used in nuclear and particle
physics. In terms of compact Lie groups it is $SU(3)\supset
U(2Á)=SU(2)\times U(1)$. The reduction is accomplished by applying
to each element of the $A_2$ orbit the projection
matrix~${\mbox P=\left(\begin{smallmatrix} 1 & 0\\ 1 & 2 \end{smallmatrix}\right)}$,
and by subsequently regrouping the results into orbits of $A_1\times
U_1$. We find the branching rules for the three types of $A_2$
orbits:
\begin{align}
(p,0)&\supset (p)(p){+}(0)(-2p),\nonumber\\
(0,q)&\supset (q)(-q){+}(0)(2q),\nonumber\\
(p,q)&\supset (p)(p{+}2q){+}(p{+}q)(p-q){+}(q)(-2p-q),\nonumber
\end{align}
where $p,q\in\R^{>0}$.

\subsubsection{$A_3\supset A_2\times U_1$}\

Reduction is achieved by applying to each element of an $A_3$ orbit the projection matrix
\begin{equation}\label{matrixA4}
\left(\begin{smallmatrix} 1 & 0 & 0 \\
                          0 & 1 & 0 \\
                          1 & 2 & 3
\end{smallmatrix}\right)
\end{equation}
and by subsequently regrouping the results into orbits of $A_2\times
U_1$. For all seven types of $A_3$ orbits, we find the branching
rules:
\begin{align}
(p,0,0)&\supset (p,0)(p){+}(0,0)({-}3p),\nonumber\\
(0,q,0)&\supset (0,q)(2q){+}(q,0)({-}2q),\nonumber\\
(0,0,r)&\supset (0,0)(3r){+}(0,r)({-}r),\nonumber\\
(p,q,0)&\supset (p,q)(p{+}2q){+}(p{+}q,0)(p{-}2q){+}(q,0)({-}3p{-}2q),\nonumber\\
(p,0,r)&\supset (p,0)(p{+}3r){+}(p,r)(p{-}r){+}(0,r)({-}3p{-}r),\label{example1}\\
(0,q,r)&\supset (0,q)(2q{+}3r){+}(0,q{+}r)(2q{-}r){+}(q,r)({-}2q{-}r),\nonumber\\
(p,q,r)&\supset (p,q)(p{+}2q{+}3r){+}(p,q{+}r)(p{+}2q{-}r) \nonumber\\
          &\quad {+}(p{+}q,r)(p{-}2q{-}r){+}(q,r)({-}3p{-}2q{-}r),\nonumber
\end{align}
where $p,q,r\in\R^{>0}$.

\begin{example}\

Let us illustrate the actual computation of branching rules on
the example of $A_3$ orbit $(2,0,1)$ containing 12 points. We write
the coordinates of the points as column vectors:
\begin{gather}\label{A4orbit}
\begin{gathered}
\left(\begin{smallmatrix} 2 \\ 0 \\ 1  \end{smallmatrix}\right),
\left(\begin{smallmatrix} \!{-}2 \\ 2 \\ 1  \end{smallmatrix}\right),
\left(\begin{smallmatrix} 2 \\ 1 \\\!{-}1  \end{smallmatrix}\right),
\left(\begin{smallmatrix} 0 \\{-}2 \\ 3  \end{smallmatrix}\right),
\left(\begin{smallmatrix}\!{-}2 \\ 3 \\\!{-}1  \end{smallmatrix}\right),
\left(\begin{smallmatrix} 3 \\{-}1 \\ 0  \end{smallmatrix}\right),
\\
\left(\begin{smallmatrix} 0 \\ 1 \\\!{-}3  \end{smallmatrix}\right),
\left(\begin{smallmatrix}1 \\\!{-}3 \\ 2  \end{smallmatrix}\right),
\left(\begin{smallmatrix}\!{-}3 \\ 2 \\ 0  \end{smallmatrix}\right),
\left(\begin{smallmatrix} 1 \\\!{-}1 \\\!{-}2  \end{smallmatrix}\right),
\left(\begin{smallmatrix} \!{-}1 \\\!{-}2 \\ 2  \end{smallmatrix}\right),
\left(\begin{smallmatrix}\!{-}1 \\ 0 \\\!{-}2  \end{smallmatrix}\right).
\end{gathered}
\end{gather}
Multiplying each of the points of \eqref{A4orbit} by the matrix
\eqref{matrixA4}, one gets the points of the $A_2\times U_1$ orbits
written as column vectors. Rewriting them in the horizontal form
and remembering that the first two coordinates belong to $A_2$
orbits, the third one belonging to $U_1$, we have the set of
projected points. It remains to distribute the points into
individual orbits. Practically it suffices to select just the
dominant ones (no negative coordinates) because they represent the
orbits that are present. Results are given by
\eqref{example1}, where $p=2$ and $r=1$.
\end{example}

\subsubsection{$A_n\supset A_{n-1}\times U_1$,\quad  $n\geq 2$}\

The cases listed in 4.1.2 and 4.1.3 are special cases of the present one.
\begin{gather*}
\left(\begin{smallmatrix} 1 & 0 & 0 & 0 & \cdots & 0 & 0 & 0 \\
                          0 & 1 & 0 & 0 & \cdots & 0 & 0 & 0 \\
                          0 & 0 & 1 & 0 & \cdots & 0 & 0 & 0 \\[-5pt]
                          \vdots & \vdots & \vdots  & \vdots  & \ddots & \vdots  &  \vdots & \vdots \\[1 pt]
                          0 & 0 & 0 & 0 & \cdots & 0 & 1 & 0 \\
                          1 & 2 & 3 & 4 & \cdots & n{-}2 & n{-}1 & n
\end{smallmatrix}\right)\,
\end{gather*}
We write here just a few branching rules for this case:
\begin{align}
(p,0,0,\cdots,0)&\supset (p,0,0,\cdots,0)(p){+}(0,\cdots,0)(-np),\nonumber\\
(0,q,0,\cdots,0)&\supset (0,q,0,\cdots,0)(2q){+}(q,0,0,\cdots,0)(-(n-1)q), \nonumber\\
(p,0,\cdots,0,r)&\supset (p,0,0,\cdots,0)(p{+}nr){+}(p,0,0,\cdots,0,r)(p-r)\nonumber\\
                &\quad {+}(0,0,\cdots,0,r)(-np-r).\nonumber
\end{align}
Note that, here and everywhere below, $p,q,r\in\R^{>0}$.

\subsection{Orbit branching rules for
$A_n\supset A_{n-k-1}\times A_k\times U_1$}\

All the cases so far can be viewed as the special cases of the
present one where ${\mbox k=0}$. Here we are considering the cases with
general rank $n\geq3$ and $1\leq~k\leq~[\tfrac{n-1}2]$.

\subsubsection{$A_3\supset A_1\times A_1\times U_1$}\

The reduction is accomplished by applying to each element of the
$A_3$ orbit the projection matrix~$P=\left(\begin{smallmatrix} 1 & 0 &
0 \\ 0 & 0 & 1 \\ 1 & 2 & 1 \end{smallmatrix}\right)$, and by
subsequently regrouping the results into orbits of $A_1\times
A_1\times U_1$.

For all types of $A_3$ orbits, we find:
\begin{align}
(p,0,0)&\supset (p)(0)(p){+}(0)(p)({-}p), \nonumber\\
(0,q,0)&\supset (0)(0)(2q){+}(0)(0)({-}2q){+}(q)(q)(0),\nonumber\\
(0,0,r)&\supset (0)(r)(r){+}(r)(0)({-}r),\nonumber\\
(p,q,0)&\supset (p)(0)(p{+}2q){+}(p{+}q)(q)(p){+}(q)(p{+}q)({-}p){+}(0)(p)({-}p{-}2q),\nonumber\\
(p,0,r)&\supset (p)(r)(p{+}r){+}(p{+}r)(0)(p{-}r){+}(0)(p{+}r)(r{-}p){+}(r)(p)({-}p{-}r),\nonumber\\
(0,q,r)&\supset (0)(r)(2q{+}r){+}(q)(q{+}r)(r){+}(q{+}r)(q)({-}r){+}(r)(0)({-}2q{-}r),\nonumber\\
(p,q,r)&\supset (p)(r)(p{+}2q{+}r){+}(p{+}q)(r{+}q)(p{+}r){+}(p{+}q{+}r)(q)(p{-}r)\nonumber\\
            &\quad {+}(q)(p{+}q{+}r)(r{-}p){+}(q{+}r)(p{+}q)({-}p{-}r){+}(r)(p)({-}p{-}2q{-}r).\nonumber
\end{align}

\subsubsection{$A_4\supset A_2\times A_1\times U_1$}\

In terms of compact Lie groups, this is the case frequently used in
particle physics, namely $SU(5)\supset SU(3)\times SU(2)\times
U(1)$. The reduction is accomplished by applying to each element of
the $A_4$ orbit the projection matrix
\begin{gather*}
\left(\begin{smallmatrix} 1 & 0 & 0 & 0 \\
                          0 & 1 & 0 & 0 \\
                          0 & 0 & 0 & 1 \\
                          2 & 4 & 6 & 3
\end{smallmatrix}\right)
\end{gather*}
and by subsequently regrouping the results into orbits of $A_2\times
A_1\times U_1$. For the following types of $A_4$ orbits, we find:
\begin{align*}
(p,0,0,0)&\supset (p,0)(0)(2p){+}(0,0)(p)(-3p),\nonumber \\
(0,q,0,0)&\supset (0,q)(0)(4q){+}(q,0)(q)(-q){+}(0,0)(0)(-6q), \nonumber\\
(p,0,0,r)&\supset (p,0)(r)(2p{+}3r){+}(p,r)(0)(2p-2r)\nonumber\\
         &\quad{+}(0,0)(p{+}r)(3r-3p){+}(0,r)(p)(-3p-2r).\nonumber
\end{align*}

\subsubsection{$A_n\supset A_{n-2}\times A_1\times U_1$, for odd $n\geq 3$}\

The projection matrix is
\begin{gather*}
\left(\begin{smallmatrix}
1 & 0 & 0 & 0 & \cdots & 0 & 0 & 0 & 0 \\
0 & 1 & 0 & 0 & \cdots & 0 & 0 & 0 & 0 \\
0 & 0 & 1 & 0 & \cdots & 0 & 0 & 0 & 0 \\[-5pt]
\vdots & \vdots & \vdots  & \vdots  & \ddots & \vdots  & \vdots  & \vdots & \vdots \\[1pt]
0 & 0 & 0 & 0 & \cdots & 0 & 1 & 0 & 0 \\
0 & 0 & 0 & 0 & \cdots & 0 & 0 & 0 & 1 \\
1 & 2 & 3 & 4 & \cdots & n{-}3 & n{-}2 & n{-}1 & \tfrac{n-1}{2}
\end{smallmatrix}\right),
\end{gather*}
and some of the branching rules are
\begin{align*}
(p,0,0,\cdots,0)&\supset (p,0,\cdots,0)(0)(p){+}(0,\cdots,0)(p)({-}\tfrac{n{-}1}{2}p), \\
(0,q,0,\cdots,0)&\supset (0,q,0,\cdots,0)(0)(2q){+}(q,0,0,\cdots,0)(q)({-}\tfrac{n{-}3}{2}q)\\
                &\quad{+}(0,\cdots,0)(0)((1{-}n)q), \\
(p,0,\cdots,0,r)&\supset (p,0,\cdots,0)(r)(p{+}\tfrac{n{-}1}{2}r){+}(p,0,\cdots,0,r)(0)(p{-}r)\\
                &\quad{+}(0,\cdots,0)(p{+}r)((r{{-}}p)\tfrac{n{-}1}{2}){+}(0,\cdots,0,r)(p)({-}r{-}\tfrac{n{-}1}{2}p).
\end{align*}

\subsubsection{$A_n\supset A_{n-2}\times A_1\times U_1$,\quad for even $n\geq 4$}\

The projection matrix is
\begin{gather*}
\left(\begin{smallmatrix}
1 & 0 & 0 & 0 & \cdots & 0 & 0 & 0 & 0 \\
0 & 1 & 0 & 0 & \cdots & 0 & 0 & 0 & 0 \\
0 & 0 & 1 & 0 & \cdots & 0 & 0 & 0 & 0\\[-5pt]
\vdots & \vdots & \vdots  & \vdots  & \ddots & \vdots  & \vdots  & \vdots & \vdots \\[1pt]
0 & 0 & 0 & 0 & \cdots & 0 & 1 & 0 & 0 \\
0 & 0 & 0 & 0 & \cdots & 0 & 0 & 0 & 1 \\
2 & 4 & 6 & 8 & \cdots & 2(n{-}3) & 2(n{-}2) & 2(n{-}1) & n{-}1
\end{smallmatrix}\right),
\end{gather*}
and some of the branching rules are
\begin{align*}
(p,0,0,\cdots,0)&\supset (p,0,\cdots,0)(0)(2p){+}(0,\cdots,0)(p)((1-n)p), \\
(0,q,0,\cdots,0)&\supset (0,q,0,\cdots,0)(0)(4q){+}(q,0,\cdots,0)(q)((3-n)q)\\
                &\quad{+}(0,\cdots,0)(0)(2(1-n)q), \\
(p,0,\cdots,0,r)&\supset (p,0,\cdots,0)(r)(2p{+}(n-1)r){+}(p,0,\cdots,0,r)(0)(2(p-r))\\
                &\ {+}(0,\cdots,0)(p{+}r)((n-1)(r-p)){+}(0,\cdots,0,r)(p)(-2r-(n-1)p).
\end{align*}

\subsection{The general case $A_{n}\supset A_{n-k-1}\times A_{k}\times U_1\colon$ $1\leq~k\leq~[\tfrac{n-1}2]$}\

The branching rules of subsection 4.2 are important
special cases of the general case. The projection matrix in the general case can be written as:

\begin{gather*}
\left(
\begin{array}{ccc}
{}&\multicolumn{2}{|c}{}\\
I_{n-k-1}&\multicolumn{2}{|c}{{\bf 0}}\\[1ex]
\hline
\multicolumn{2}{c|}{ }&{}\\[0.1 ex]
\multicolumn{2}{c|}{{\bf 0}}&I_{k}\\[1 ex]
\hline
{}&{}&{}\\[-4 pt]
\text{{\tiny$k\!{+}\!1$}}\;
\text{{\tiny$2(k\!{+}\!1)$}}
\cdots
\text{{\tiny$(n\!{-}\!k\!{-}\!1)(k\!{+}\!1)$}}
&
\text{{\tiny$(n\!{-}\!k)(k\!{+}\!1)$}}\!\!
&
\text{{\tiny$k(n\!{-}\!k)$}}
\cdots
\text{{\tiny$2(n\!{-}\!k)$}}\;
\text{{\tiny$n\!{-}\!k$}}
\\[-2 ex]
\underbrace{\qquad\qquad\qquad\qquad\;\;\;}_{n-k-1}
&\underbrace{\qquad\;\;\;\;}_{1}
&\underbrace{\qquad\qquad\qquad\quad}_{k}
\end{array}
\right)
\!\!\!\!\!\!\!\!\!\!\!
\begin{array}{l}
\\[-1 ex]
\left.
\begin{array}{l}
{}\\
{}\\[0,5 ex]
\end{array}
\right\}\text{\tiny $n{-}k{-}1$}\\
\\[-3 ex]
\left.
\begin{array}{l}
{}\\
{}\\[1.5 ex]
\end{array}
\right\}\text{\tiny $k$}\\
\\[-2,8ex]
\left.
\begin{array}{l}
{}\\
\end{array}
\right\}\text{\tiny $1$}\\[5ex]
\end{array}
\end{gather*}
Note that, here and everywhere below, $I_k$
denotes the $k\times k$ identity matrix and ${\bf 0}$ represents the zero matrix.

\begin{align}
(p,0,0,\dots,0)&{\supset}(p,0,\dots,0)(0,\dots,0)((k{+}1)p)
                   {+}(0,\dots,0)(p,0,\dots,0)((k{-}n)p),
                   \notag\\
(p,0,\cdots,0,r)&{\supset}(p,0,\cdots,0)(0,\dots,0,r)((k{+}1)p{+}(n-k)r) \notag\\
                &\phantom{\supset}{+}(p,0,\cdots,0,r)(0,\cdots,0)((k{+}1)(p-r)) \notag\\
                &\phantom{\supset}{+}(0,\cdots,0)(p,0,\cdots,0,r)((n-k)(r-p)) \notag\\
                &\phantom{\supset}{+}(0,\cdots,0,r)(p,0,\cdots,0)((-k-1)r{+}(k-n)p).\notag
\end{align}

\section{Branching rules for maximal semisimple subalgebras of $A_n$}

The simple Lie algebras $A_n$ contain no semisimple subalgebras of
the same rank $n$. Hence all orbit branching rules considered
in this Section have rank strictly smaller than $n$. We
proceed by increasing rank values until $n=8$. Then we describe
the infinite series involving the Weyl groups of classical Lie
algebras, namely $W(A_{2n})\supset W(B_n),\  n\geq3$,
$W(A_{2n-1})\supset W(C_n),\  n\geq2$, and $W(A_{2n-1})\supset
W(D_n),\  n\geq4$.

We include the low-rank special cases of the three infinite
series. We exclude the cases when a subalgebra is maximal
among semisimple Lie algebras, but not among reductive algebras.
Projection matrices for the latter cases are obtained by striking
the last line of the corresponding matrices from the previous Section.
Dots in the projection matrices denote zero matrix elements.

\subsection{Rank 2}\

There is only one case here, namely $A_2\supset A_1$, which is often
specified in terms of corresponding Lie groups either as
$SU(3)\supset O(3)$, if the groups should be compact, or
$Sl(3,\C)\supset O(3,\C)$, if the groups have complex parameters.
Their Weyl group orbits are the same. The projection matrix is
$P=(2\ 2)$, so that we obtain the reductions:
\begin{gather} \label{BRa2}
(p,q) \supset (2p{+}2q){+}(2p){+}(2q), \quad (p,0) \supset
(2p){+}(0),\quad (0,q) \supset  (2q){+}(0).
\end{gather}

\begin{example}\

Let us underline the geometrical content of the relations \eqref{BRa2}.
On the left side, there are points in $\R^2$ given by their
coordinates in $\omega$-basis $\{\omega_1,\ \omega_2\}$ of $A_2$.
The geometric relation of the two basis vectors is given by the
$2\times2$ matrix of scalar products $\l\omega_j,\omega_k\r$. In
$A_n$, it happens to be the inverse $C^{-1}$ of the Cartan matrix
of the algebra. In particular, for $A_2$, we have
$C^{-1}=\tfrac13\left(\begin{smallmatrix}2&1\\1&2\end{smallmatrix}\right)$.
It follows that the basis vectors are of equal length $\sqrt{2/3}$, and
that $\angle (\omega_1,\omega_2)=60^\circ$.

On the right side of \eqref{BRa2}, there are the $A_1$ orbit points
in $\R^1\subset\R^2$. Applying to $A_1$ the same rules as previously
applied to $A_2$, we have $C=(2)$ so that $C^{-1}=(1/2)$. Thus the
basis vector of $A_1$, say $\omega$, has the length $1/\sqrt2$.

It remains to clarify what are the relative positions of
$\omega_1,\omega_2$, and $\omega$. The theory leaves us several
options. A reasonable choice is built-in into the construction of
the projection matrix in each case. Justification for this
is outside the scope of this paper. However, the relative positions
of basis vectors in $\R^2$ and $\R^1$ are established, for example, from
\begin{gather*}
P\omega_1=(2\ 2)\left(\begin{smallmatrix}1\\0\end{smallmatrix}\right)=P\omega_2=2\omega.
\end{gather*}
Since equal-length vectors $\omega_1$ and $\omega_2$ are projected
into the same point on the $\omega$-axis, the direction of $\omega$
divides the angle between $\omega_1$ and $\omega_2$ into equal
parts.
\end{example}\

\subsection{Rank 3}\

There are just two cases to consider. We write only their projection matrices.
\begin{gather*}
A_3\supset C_2:\quad
   \left( \begin{smallmatrix} 1 & \c & 1 \\ \c & 1 & \c\end{smallmatrix}\right),\qquad
A_3\supset A_1\times A_1:\quad
   \left( \begin{smallmatrix} 1 & \c & 1 \\ 1 & 2 & 1\end{smallmatrix}\right).
\end{gather*}

\begin{example}\

There are 12 points in \eqref{A4orbit}. Let us transform them by the
matrix $\left( \begin{smallmatrix} 1 & \c & 1 \\ \c & 1 &
\c\end{smallmatrix}\right)$. Two dominant points are found
when writing the projected points in horizontal form, namely
$(3,0)$ and $(1,1)$. Hence we have the $A_3{\supset}C_2$ rule
${\mbox (2,0,1){\supset}(3,0){+}(1,1)}$. The orbit $(3,0)$ contains 4 points
and the orbit $(1,1)$ contains 8 points.

Geometrically, $(2,0,1)$ is a tetrahedron with four cut-off
vertices. The planar figure after the projection is the union of
the square $(3,0)$ and the octagon $(1,1)$.
\end{example}

Let us underline the difference between the subalgebra $A_1\times
A_1$ here and the one in subsection 4.2.1. Using the corresponding
projection matrices, we obtain respectively the reductions
\begin{gather*}
(1,0,0)\supset(1)(1),\quad\text{and}\quad (1,0,0)\supset
(1)(0)(1){+}(0)(1)(-1).
\end{gather*}
Ignoring the contribution from $U_1$ in the second branching rule,
the four orbit points  obtained after the reduction are different in
the two cases:
\begin{align*}
(1,0,0)&\supset\{(1)(1),\ (-1)(1),\ (1)(-1),\ (-1)(-1)\},\\
(1,0,0)&\supset\{(1)(0),\ (-1)(0),\ (0)(1),\ (0)(-1)\}.
\end{align*}

There is an obvious subalgebra $A_2$ in $A_3$. Although it is
maximal among semisimple subalgebras of $A_3$, it is not maximal
among reductive subalgebras. It coincides with $A_2$ in
subsection 4.1.3.


\subsection{Rank 4}\

There is only one simple and maximal subalgebra of $A_4$
among the reductive subalgebras:
\begin{gather*}
A_4\supset C_2:\quad
   \left(\begin{smallmatrix}\c & 2 & 2 & \c \\
                             1 &\c &\c & 1
         \end{smallmatrix}\right).
\end{gather*}
The other two semisimple subalgebras of rank 3 of $A_4$, namely
$A_3$ and $A_1\times A_2$, can be both extended by $U_1$ to maximal
reductive subalgebras. They are the special cases $n=4$ found in
subsections 4.1.4 and 4.2.4 respectively.

Some branching rules:
\begin{align*}
(p,0,0,0)&\supset(0,p){+}(0,0),\\
(p,0,0,r)&\supset(0,p{+}r){+}(0,p){+}(0,r){+}(2r,p-r), \quad p>r,\\
(p,0,0,p)&\supset(0,2p){+}2(0,p){+}2(2p,0).
\end{align*}

\subsection{Rank 5}\

There are four maximal subalgebras in this case. The first two are
special cases of the general inclusions of subsection 5.8. The Lie
algebras $A_3$ and $D_3$ coincide, except that by general convention
we agreed not to consider the $D_3$ form. Therefore $A_5\supset
A_3$ can be read equivalently as $A_5\supset D_3$, provided that we
modify the order of point coordinates as follows: $(a,b,c)$
of $A_3$ corresponds to $(b,a,c)$ of $D_3$.
\begin{alignat*}{3}
&A_5\supset A_3:\quad
   \left(\begin{smallmatrix} \c & 1 & \c & 1 & \c \\1 & \c & \c & \c & 1 \\
                             \c & 1 & 2 & 1 & \c
         \end{smallmatrix}\right),\qquad
&&A_5\supset C_3:  \quad
   \left(\begin{smallmatrix} 1 & \c & \c & \c & 1 \\ \c & 1 & \c & 1 & \c \\
                             \c & \c & 1 & \c & \c
         \end{smallmatrix}\right),\\
&A_5\supset A_2:  \quad
   \left(\begin{smallmatrix} \c & 1 & 3 & 2 & 2 \\ 2 & 2 & \c & 1 & \c
         \end{smallmatrix}\right),\qquad
&&A_5\supset A_1\times A_2: \quad
   \left(\begin{smallmatrix} 1 & \c & 1 & \c & 1  \\ 1 & 2 & 1 & \c & \c \\
                             \c & \c & 1 & 2 & 1
         \end{smallmatrix}\right).
\end{alignat*}

In particular, the branching rules for the $A_5$ orbit of 6 points are:
\begin{align}
(p,0,0,0,0)\supset
   \begin{cases} (0,p,0)\quad &\text{for}\quad A_3\\
                 (p,0,0)\quad &\text{for}\quad C_3\\
                 (0,2p)\quad &\text{for}\quad C_2\\
                 (p)(p,0)\quad &\text{for}\quad A_1\times A_2\\
   \end{cases}\qquad p\in\R^{>0}.
\end{align}
The first two are special cases of~(\ref{A-Dn}) and (\ref{A-Cn}) respectively.

\subsection{Rank 6}\

The only entry here is a special case of $A_{2n}\supset B_n$ of
subsection 5.8, and its branching rules.
\begin{equation*}
A_6\supset B_3 : \quad\left(
\begin{smallmatrix} 1 & \c & \c & \c & \c & 1 \\
                   \c & 1 & \c & \c & 1 & \c \\
                   \c & \c & 2 & 2 & \c & \c
\end{smallmatrix}\right).
\end{equation*}

\begin{align*}
(p,0,0,0,0,0)&\supset(p,0,0){+}(0,0,0),\\
(p,0,0,0,0,r)&\supset(p{+}r,0,0){+}(p,0,0){+}(r,0,0){+}(p-r,r,0),\quad p>r,\\
(p,0,0,0,0,p)&\supset(2p,0,0){+}2(p,0,0){+}2(0,p,0).
\end{align*}


\subsection{Rank 7}\

The first two of the three cases are restrictions to $n=7$ of
the corresponding general cases of subsection 5.8.
\begin{gather*}
A_7\supset C_4 :\quad \left(
\begin{smallmatrix}1 & \c & \c & \c & \c & \c & 1 \\
                   \c & 1 & \c & \c & \c & 1 & \c \\
                   \c & \c & 1 & \c & 1 & \c & \c \\
                   \c & \c & \c & 1 & \c & \c & \c
\end{smallmatrix}\right),\qquad
A_7\supset D_4 :\quad\left(
\begin{smallmatrix}1 & \c & \c & \c & \c & \c & 1 \\
                   \c & 1 & \c & \c & \c & 1 & \c \\
                   \c & \c & 1 & \c & 1 & \c & \c \\
                   \c & \c & 1 & 2 & 1 & \c & \c
\end{smallmatrix}\right),\\
A_7\supset A_1\times A_3 :\quad \left(
\begin{smallmatrix}1 & \c & 1 & \c & 1 & \c & 1 \\
                   1 & 2 & 1 & \c & \c & \c & \c \\
                   \c & \c & 1 & 2 & 1 & \c & \c \\
                   \c & \c & \c & \c & 1 & 2 & 1
\end{smallmatrix}\right).
\end{gather*}

In particular, for $A_7\supset A_1\times A_3$, we obtain
\begin{align*}
(p,0,0,0,0,0,0)&\supset (p)(p,0,0), \\
(p,0,0,0,0,0,r)&\supset (p{+}r)(p,0,r){+}(p{-}r)(p,0,r){+}(p{+}r)(p{-}r,0,0), \quad p>r,\\
(p,0,0,0,0,0,p)&\supset (2p)(p,0,p){+}2(0)(p,0,p){+}4(2p)(0,0,0).
\end{align*}

\subsection{Rank 8}\

The first case is a special case of~(\ref{A-Bn}).
\begin{gather*}
A_8\supset B_4 :\quad \left(
\begin{smallmatrix}1 & \c & \c & \c & \c & \c & \c & 1 \\
                   \c & 1 & \c & \c & \c & \c & 1 & \c \\
                   \c & \c &1 & \c & \c & 1 & \c & \c \\
                   \c & \c & \c & 2 & 2 & \c & \c & \c
\end{smallmatrix}\right),\qquad
A_8\supset A_2\times A_2 :\quad\left(
\begin{smallmatrix}1 & \c & 1 & 1 & \c & 1 & 1 & \c \\
                   \c & 1 & 1 & \c & 1 & 1 & \c & 1 \\
                   1 & 2 & 1 & 2 & 1 & 1 & \c & \c \\
                   \c & \c & 1 & 1 & 2 & 1 & 2 & 1
\end{smallmatrix}\right).
\end{gather*}

Examples of the branching rules for the second case:
\begin{align*}
(p,0,0,0,0,0,0,0)&\supset(p,0)(p,0),\\
(p,0,0,0,0,0,0,r)&\supset(p,r)(p,r){+}(p{-}r,0)(p,r){+}(p,r)(p{-}r,0), \quad p>r,\\
(p,0,0,0,0,0,0,p)&\supset(p,p)(p,p){+}3(0,0)(p,p){+}3(p,p)(0,0).
\end{align*}

\subsection{Three general rank cases}\

The cases are presented with examples of branching rules for the
orbits $(p,0,\dots,0)$ and $(p,0,\dots,0,r)$, where the parameters
$p,\,r$ are strictly positive and real. We also assume that $p>r$.
If $p<r$ the parameters $p$ and $r$ in the branching rule need to be
interchanged. The case $p=r$ often needs to be listed separately.

\medskip
$A_{2n}\supset B_n$,\quad $n\geq 3$
\begin{gather}\label{A-Bn}
P=\left(\!\!
\begin{array}{c|c|c}
{}&{}&{}\\[-1,5ex]
I_{n-1}&{\bf 0}&E_{n-1}\\[0,5 ex]
\hline 0 \cdots 0 & 2\; 2& 0 \cdots 0
\end{array}
\!\!\right),
\end{gather}
Note that, here and everywhere below, $E_k$
denotes the $k\times k$ matrix with units on the codiagonal.
\begin{align*}
(p,0,0,\cdots,0)&{\supset} (p,0,\cdots,0){+}(0,\cdots, 0), \\
(p,0,\cdots,0,r)&{\supset} (p{+}r,0,\cdots,0){+}(p,0,\cdots,0){+}(r,0,\cdots,0){+}(p{-}r,r,0,\cdots,0),\\
(p,0,\cdots,0,p)&{\supset}
(2p,0,\cdots,0){+}2(p,0,\cdots,0){+}2(0,p,0,\cdots,0).
\end{align*}

\medskip

$A_{2n-1}\supset C_n$,\quad $n\geq 2$
\begin{gather}\label{A-Cn}
P=\left(\!\!
\begin{array}{c|c|c}
{}&{}&{}\\[-1,5ex]
I_{n-1}&{\bf 0}&E_{n-1}\\[0,5 ex]
\hline 0 \cdots 0 & 1& 0 \cdots 0
\end{array}
\!\!\right)
\end{gather}
\begin{align*}
(p,0,0,\cdots,0)&\supset (p,0,\cdots,0),\\
(p,0,\cdots,0,r)&\supset (p{+}r,0,\cdots,0){+}(p-r,r,0,\cdots,0),\\
(p,0,\cdots,0,p)&\supset (2p,0,\cdots,0){+}2(0,p,0,\cdots,0).
\end{align*}

\medskip

$A_{2n-1}\supset D_n$,\quad $n\geq 4$
\begin{gather}\label{A-Dn}
P=\left(\!\!
\begin{array}{c|c|c}
{}&{}&{}\\[-1,5ex]
I_{n-1}&{\bf 0}&E_{n-1}\\[0,5 ex]
\hline 0 \cdots 0\  1 & 2& 1\ 0 \cdots 0
\end{array}
\!\!\right)
\end{gather}
\begin{align*}
(p,0,0,\cdots,0)&\supset (p,0,\cdots,0),\\
(p,0,\cdots,0,r)&\supset (p{+}r,0,\cdots,0){+}(p-r,r,0,\cdots,0), \\
(p,0,\cdots,0,p)&\supset (2p,0,\cdots,0){+}2(0,p,0,\cdots,0).
\end{align*}

\section{Concluding remarks}
\begin{itemize}
\item
\smallskip
The pairs  $W(L)\supset W(L')$ in this paper involve a maximal
subalgebra $L'$ in~$L$. A chain of maximal subalgebras linking $L$
and any of its reductive non-maximal subalgebras $L''$ can be found.
Corresponding projection matrices combine, by the common matrix
multiplication, into the projection matrix for $W(L)\supset
W(L'')$.
\smallskip

\item
Projection matrices of Section~4 are square matrices with determinant
different from zero. Hence they can be inverted and used in the opposite
direction. The inverse matrix transforms an orbit of $W(L')$ into the linear combination of orbits of $W(L)$, where $L'\subset L$. The linear combination has integer coefficients of both signs in general. We know of no interpretation of such `branching rules' in applied literature, although they have their place in the Grothendieck rings of representations.
\smallskip

\item
Curious and completely unexplored relations between pairs of maximal
subalgebras, say $L'$ and $L''$, of the same Lie algebra $L$ can be
found by combining the projection matrices $P(L\rightarrow L')$ and
$P(L\rightarrow L'')$ as
\begin{gather*}
P(L'\rightarrow L'')=P(L\rightarrow L'')P^{-1}(L\rightarrow L').
\end{gather*}

\item
The index of a semisimple subalgebra in a simple Lie algebra is an
invariant of all branching rules for a fixed algebra-subalgebra
pair. It was introduced in \cite{Dynkin}, see Equation (2.26). It is an
invariant also for any pair $W(L)\supset W(L')$.
\smallskip

\item
Congruence classes of representations are naturally extended to
congruence classes of $W$-orbits \cite{HLP}. Comparing the
congruence classes of orbits for $W(L)\supset W(L')$ reveals that
not all combinations of congruence classes are present. A
relative congruence class is a valid and useful concept which
deserves investigation.
\smallskip

\item
Here, the relations between orbits were defined by the
classification of maximal reductive subalgebras in simple type $A_n$ Lie
algebras. There exists another relation between such
algebras that is not an homomorphism. It is called subjoining
\cite{MPi,PSS}. Consider an example. The 4-dimensional
representation $(1,0,0)$ of $A_3$ does {\it not} contain the 5-dimensional representation $(0,1)$ of $C_2$. In spite of that,
the projection matrix that maps the highest weight orbit of $A_3$ to the orbit $(0,1)$ of $C_2$ can be obtained.
Indeed, that projection matrix is
$\left(\begin{smallmatrix}0&2&0\\1&0&1\end{smallmatrix}\right)$.

\end{itemize}
\medskip
\medskip
\centerline{\bf Acknowledgements}
\medskip

This work was supported in part by the Natural Sciences and Engineering
Research Council of Canada and by the MIND Research Institute. We are
grateful for the hospitality extended to M.~N. at the
Centre de recherches math\'ematiques, Universit\'e de Montr\'eal,
where the work was done.


\begin{thebibliography}{14}
\footnotesize

\bibitem{BdeS}
A. Borel, J. de Siebental, {\it Les sous-groupes ferm\'es de rang
maximum de groupes de Lie clos,\/} Comment. Math. Helv. {\bf 23}
(1949) 200--221.


\bibitem{BMP}
M.R.~Bremner, R.~V.~Moody, J.~Patera, {\sl Tables of dominant weight
multiplicities for representations of simple Lie algebras,\/} Marcel
Dekker, New York, 1985.

\bibitem{CKPS}
B.~Champagne, M.~Kjiri, J.~Patera,  R.~T.~Sharp, {\it  Description
of reflection generated polytopes using decorated Coxeter
diagrams,\/} Can. J.~Phys., {\bf 73} (1995) 566--584.

\bibitem{Dynkin}
E.B. Dynkin, {\it Semisimple subalgebras of semisimple Lie
algebras,\/} AMS Trnanslations, Series 2, Vol. 6, (1957) 111--244.

\bibitem{GPS}
F.~Gingras, J.~Patera, R.~T.~Sharp, {\it  Orbit-orbit branching
rules between simple low-rank algebras and equal-rank
subalgebras,\/} J.~Math.~Phys. {\bf 33} (1992) 1618--1626.

\bibitem{GP}
S. Grimm and J. Patera, {\it  Decomposition of tensor products of
the fundamental representations of $E_8$,\/} in {\sl Advances in
Mathematical Sciences -- CRM's 25 Years,\/} ed.~L.~Vinet, CRM Proc.
Lecture Notes, vol. 11, Amer. Math. Soc., Providence, RI, (1997)
329--355.

\bibitem{HLP}
L. H\'akov\'a, M. Larouche, J. Patera, {\it The rings of
$n$-dimensional polytopes,\/}  J.~Phys.~A: Math. Theor., {\bf 41}
(2008) 495202.

\bibitem{H1}
J.E.~Humphreys, {\sl Introduction to Lie Algebras and Representation
Theory,\/} Springer, New York, 1972.

\bibitem{H}
J.E.~Humphreys, {\sl Reflection Groups and Coxeter Groups,\/}
Cambridge Univ. Press, Cambridge, 1990.



\bibitem{KP1}
Klimyk~A. and Patera~J., Orbit functions \textit{SIGMA} \textbf{2}
paper~006, 60~pages, math-ph/0601037.

\bibitem{KP2}
Klimyk~A. and Patera~J., Antisymmetric orbit functions
\textit{SIGMA} \textbf{3} paper~023, 83~pages, math-ph/0702040v1.

\bibitem{McP}
W. G. McKay, J. Patera, {\sl Tables of dimensions, indices, and
branching rules for representations of simple Lie algebras}, Marcel
Dekker, New York, 1981.

\bibitem{MPR}
W.~G.~McKay, J.~Patera, D.~Rand, {\it  Tables of representations of
simple Lie algebras, Vol. I: Exceptional simple Lie algebras,\/} Les
Publications CRM, Montr\'eal, 1990, 318 pages.

\bibitem{MPS}
W.~G.~McKay, J.~Patera, D.~Sankoff, {\it The computation of
branching rules for representations of semisimple Lie algebras,\/}
in Computers in Nonassociative Rings and Algebras,  ed. J. Beck and
B. Kolman, Academic Press, New York, 1977.

\bibitem{MP}
R. V.~Moody, J.~Patera, {\it Characters of elements of finite order
in simple Lie groups,\/} SIAM J. on Algebraic and Discrete Methods
{\bf 5} (1984) 359--383.

\bibitem{MP1}
R. V. Moody, J. Patera, {\it  Computation of character
decompositions of class  functions on compact semisimple Lie
groups,\/} Mathematics of Computation {\bf 48} (1987) 799--827.

\bibitem{MPi}
R.V. Moody, A. Pianzola, {\it $\lambda$-mappings of representation
rings of Lie algebras,\/} Can. J. Math. {\bf 35} (1983) 898--960.

\bibitem{NPT}
M. Nesterenko, J. Patera, A. Tereszkjewicz {\it Orbit functions of
$SU(n)$ and Chebyshev polynomials,\/} 2009, arXiv:0905.2925v1.

\bibitem{P}
J. Patera, Compact simple Lie groups and theirs $C$-, $S$-, and
$E$-transforms, {\it SIGMA} {\bf 1} (2005), 025, 6 pages,
math-ph/0512029.

\bibitem{PSan}
J.~Patera, D.~Sankoff, {\it Branching rules for representations of
simple Lie algebras,\/} Presses Universit\'e de Montr\'eal,
Montr\'eal, 1973, 99 pages.

\bibitem{PSS}
J.~Patera, R.~T.~Sharp, R.~Slansky, {\it On a new relation between
semisimple Lie algebras,\/}  J.~Math.~Phys., {\bf 21} (1980)
2335--2341.

\bibitem{ST1}
M. Thoma and R. T. Sharp, {\it Orbit-orbit branching rules between
classical simple Lie algebras and maximal reductive subalgebras,\/}
J. Math. Phys., {\bf 37} (1996) 6570--6581.

\bibitem{ST2}
M. Thoma and R. T. Sharp, {\it Orbit-orbit branching rules for
families of classical Lie algebra-subalgebra pairs,\/} J. Math.
Phys., {\bf 37} (1996) 4750--4757.

\end{thebibliography}
\end{document}